\documentclass[aps,prl,reprint,superscriptaddress,showpacs]{revtex4-1} 
\usepackage{graphicx}  
\usepackage{dcolumn}   
\usepackage{amssymb}
\usepackage{amsmath}
\usepackage{physics}
\usepackage{natbib}
\usepackage{color}

\begin{document}

\title{Conductance-matrix symmetries of a three-terminal hybrid device}
\author{G.~C.~M\'enard}
\affiliation{Center for Quantum Devices, Niels Bohr Institute, University of Copenhagen, Universitetsparken 5, 2100 Copenhagen, Denmark}
\affiliation{Microsoft Quantum - Copenhagen, University of Copenhagen, Universitetsparken 5, 2100 Copenhagen, Denmark}
\author{G.~L.~R.~Anselmetti}
\affiliation{Center for Quantum Devices, Niels Bohr Institute, University of Copenhagen, Universitetsparken 5, 2100 Copenhagen, Denmark}
\affiliation{Microsoft Quantum - Copenhagen, University of Copenhagen, Universitetsparken 5, 2100 Copenhagen, Denmark}
\author{E.~A.~Martinez}
\affiliation{Center for Quantum Devices, Niels Bohr Institute, University of Copenhagen, Universitetsparken 5, 2100 Copenhagen, Denmark}
\affiliation{Microsoft Quantum - Copenhagen, University of Copenhagen, Universitetsparken 5, 2100 Copenhagen, Denmark}
\author{D.~Puglia}
\affiliation{Center for Quantum Devices, Niels Bohr Institute, University of Copenhagen, Universitetsparken 5, 2100 Copenhagen, Denmark}
\affiliation{Microsoft Quantum - Copenhagen, University of Copenhagen, Universitetsparken 5, 2100 Copenhagen, Denmark}
\author{F.~K.~Malinowski}
\affiliation{Center for Quantum Devices, Niels Bohr Institute, University of Copenhagen, Universitetsparken 5, 2100 Copenhagen, Denmark}
\affiliation{Microsoft Quantum - Copenhagen, University of Copenhagen, Universitetsparken 5, 2100 Copenhagen, Denmark}
\author{J.~S.~Lee}
\affiliation{Department of Electrical Engineering, University of California, Santa Barbara, California 93106, USA}
\author{S.~Choi}
\affiliation{California NanoSystems Institute, University of California, Santa Barbara, California 93106, USA}
\author{M.~Pendharkar}
\affiliation{Department of Electrical Engineering, University of California, Santa Barbara, California 93106, USA}
\author{C.~J.~Palmstr\o{}m}
\affiliation{California NanoSystems Institute, University of California, Santa Barbara, California 93106, USA}
\affiliation{Department of Electrical Engineering, University of California, Santa Barbara, California 93106, USA}
\affiliation{Materials Department, University of California, Santa Barbara, California 93106, USA}
\author{K.~Flensberg}
\affiliation{Center for Quantum Devices, Niels Bohr Institute, University of Copenhagen, Universitetsparken 5, 2100 Copenhagen, Denmark}
\author{C.~M.~Marcus}
\affiliation{Center for Quantum Devices, Niels Bohr Institute, University of Copenhagen, Universitetsparken 5, 2100 Copenhagen, Denmark}
\affiliation{Microsoft Quantum - Copenhagen, University of Copenhagen, Universitetsparken 5, 2100 Copenhagen, Denmark}
\author{L.~Casparis}
\email{Equal contribution, lucas.casparis@microsoft.com}
\affiliation{Center for Quantum Devices, Niels Bohr Institute, University of Copenhagen, Universitetsparken 5, 2100 Copenhagen, Denmark}
\affiliation{Microsoft Quantum - Copenhagen, University of Copenhagen, Universitetsparken 5, 2100 Copenhagen, Denmark}
\author{A.~P.~Higginbotham}
\email{Equal contribution, andrew.higginbotham@ist.ac.at}
\affiliation{Center for Quantum Devices, Niels Bohr Institute, University of Copenhagen, Universitetsparken 5, 2100 Copenhagen, Denmark}
\affiliation{Microsoft Quantum - Copenhagen, University of Copenhagen, Universitetsparken 5, 2100 Copenhagen, Denmark}

\begin{abstract}
We present conductance-matrix measurements of a three-terminal superconductor-semiconductor hybrid device consisting of two normal leads and one superconducting lead. 
Using a symmetry decomposition of the conductance, we find that the antisymmetric components of pairs of local and nonlocal conductances match at energies below the superconducting gap, consistent with expectations based on a non-interacting scattering matrix approach. 
Further, the local charge character of Andreev bound states is extracted from the symmetry-decomposed conductance data and is found to be similar at both ends of the device and tunable with gate voltage.
Finally, we measure the conductance matrix as a function of magnetic field and identify correlated splittings in low-energy features, demonstrating how conductance-matrix measurements can complement traditional tunneling-probe measurements in the search for Majorana zero modes.
\end{abstract}

\maketitle

Symmetry relations for quantum transport are often connected to deep physical principles, and make strong predictions for comparison with experiment.
For instance, the Onsager-Casimir relations \cite{onsager_reciprocal_1931,casimir_onsager's_1945,buttiker_symmetry_1988} arise from microscopic reversibility, and were central in early studies of quantum-coherent transport \cite{webb_observation_1985,benoit_asymmetry_1986,buttiker_four-terminal_1986}.
Later, predicted departures from these relations due to interaction effects \cite{christen_gauge-invariant_1996,spivak_signature_2004,sanchez_magnetic-field_2004}, which include bias-dependence of the effective potentials, were observed in nonlinear transport \cite{lofgren_symmetry_2004,zumbuhl_asymmetry_2006}.
The introduction of superconducting terminals results in additional symmetries, as conductance occurs via Andreev-reflection from electrons to holes, and is invariant under particle-hole conjugation \cite{andreev_thermal_1964}.
For a two-terminal normal-superconducting device, the conductance, $g(V)$, is a symmetric function of bias voltage, $V$, neglecting interaction effects.
As shown in a partner theoretical paper, for multi-terminal superconducting devices $g(V)$ need not be symmetric, although a curious relation exists between the antisymmetric components of the local and nonlocal conductances \cite{danon_nonlocal_2019}.
These predictions have, to our knowledge, not been tested.

Hybrid superconductor-semiconductor nanowire structures have recently become a topic of intense interest \cite{mourik_signatures_2012,das_zero-bias_2012,churchill_superconductor-nanowire_2013,albrecht_exponential_2016,deng_majorana_2016,zhang_quantized_2018}, motivated in part by proposals for achieving topological superconductivity and Majorana zero modes (MZM) \cite{lutchyn_majorana_2010,oreg_helical_2010}.
In two-terminal superconductor-semiconductor devices, observed asymmetries in the subgap conductance \cite{gul_ballistic_2018} have been suggested to arise from a dissipative fermionic reservoir, effectively acting as a third lead \cite{liu_role_2017}, although, as in the normal-conducting case \cite{buttiker_symmetry_1988}, bias-dependence of the self-consistent potential can also cause a deviation from symmetry \cite{chen_experimental_2017}.
Multi-terminal superconducting devices are a topic of particular interest, as they can be used for MZM \cite{dassarma_splitting_2012,fregoso_electrical_2013,stanescu_nonlocality_2014,rosdahl_andreev_2018,reeg_zero-energy_2018,lai_presence_2019,anselmetti_end-to-end_2019}, Cooper-pair splitter \cite{hofstetter_cooper_2009,Herrmann_2010}, and multi-terminal Josephson studies \cite{vanHeck_single_2014,strambini_omega-squipt_2016,meyer_nontrivial_2017,xie_topological_2017,pankratova_multi-terminal_2018}.
In multi-terminal superconducting quantum dot devices, bias asymmetries have been observed \cite{hofstetter_finite-bias_2011}, and a relationship between nonlocal conductance and the bound-state charge has been proposed \cite{schindele_nonlocal_2014,gramich_andreev_2017}.

\begin{figure}[!h]\vspace{-4mm}
    \includegraphics[width=1\columnwidth]{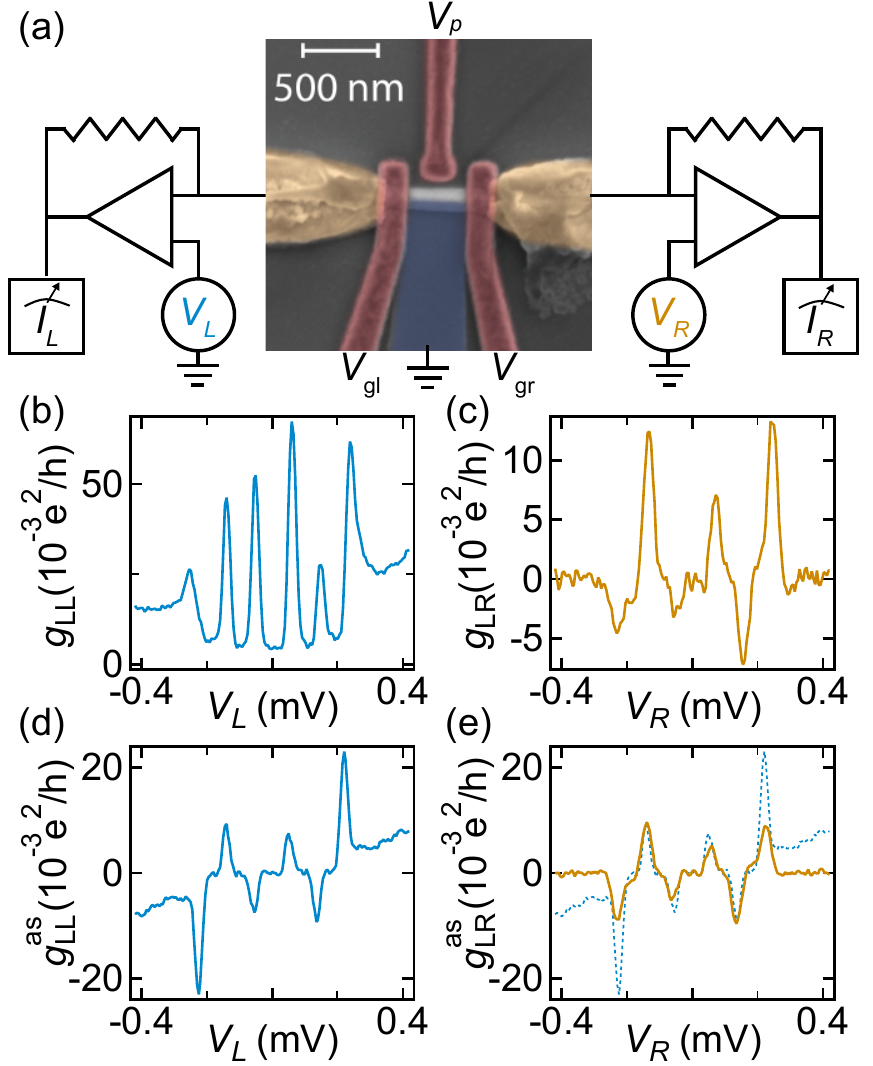}
    \caption{(a) Colored scanning electron micrograph of the three-terminal device, and schematic of the measurement setup. (b)~Left side local conductance $g_{\mathrm{LL}}$ as a function of left side bias voltage $V_L$. 
    (c)~Left side nonlocal conductance $g_{\mathrm{LR}}$ as a function of right side bias voltage $V_R$. 
    (d)~Extracted antisymmetric component of the local conductance $g_\mathrm{LL}^{\mathrm{as}}$ with respect to $V_L$. 
    (e)~Extracted antisymmetric component of the nonlocal conductance $g_\mathrm{LR}^{\mathrm{as}}$ with respect to $V_R$. The blue dashed line shows $g_\mathrm{LL}^{\mathrm{as}}$ for comparison.}
    \label{device}
\end{figure} 

In this Letter, we report a symmetry analysis of the conductance matrix measured in a three-terminal, superconductor-semiconductor hybrid device. 
The antisymmetric components for pairs of conductance matrix elements are found to coincide at energies below the superconducting gap. 
We use the symmetry-decomposed nonlocal conductance to extract the local charge character of states within the superconducting gap as a function of gate voltage, discovering that the charge is approximately equal on both sides of the device. 
Finally, we measure local and nonlocal conductances at nonzero magnetic field and identify isolated low-energy states with correlated splittings on each end of the wire, using inferred charge as an additional spectroscopic tool for comparison with theory. 
This work provides new methods for studying the local charge density of subgap bound states and distinguishing between topological and trivial states.
More generally, it provides a better understanding of the role of symmetry in multi-terminal superconducting quantum devices. 

Selective area growth (SAG) of InAs nanowires \cite{krizek_field_2018,vaitiekenas_selective-area-grown_2018} is preformed by chemical beam epitaxy (CBE) on an InP substrate masked with silicon oxide~\cite{lee_2018}. 
The nanowire is half-covered by an Al film, which was deposited in-situ after CBE growth, and is selectively etched to form a superconducting lead.
The device consists of two normal Ti/Au electrodes (yellow), a central semiconducting region proximitized by Al (blue), a global HfOx dielectric layer, and Ti/Au electrostatic gates (red). 
We emphasize that the superconducting lead is deposited during growth and contacted remote from the delicate superconductor-semiconductor interface, a benefit of the SAG approach.

Conductance is measured by applying a DC bias voltage $V_{\mathrm{bias}} = V_{R(L)}$ and an AC voltage $\delta V_{R(L)}$ at the right (left) terminal with two different AC excitation frequencies $f_R\sim$~18~Hz and $f_L\sim$~42~Hz. 
The in-phase AC current $\delta I_{R(L)}$ flowing to the right (left) side is measured with the middle superconducting lead grounded. 
The device is tunnel-coupled to the normal leads by adjusting the two tunnel-gate voltages $V_{\mathrm{gr}}$ and $V_{\mathrm{gl}}$ such that $g\ll$~e\textsuperscript{2}/h, which also ensures that the applied voltages drop over the tunnel barriers. 
We have checked for spurious voltage divider effects using a four-probe measurement on a cold-grounded device, and do not find deviations from the data presented here.
The plunger gate voltage $V_p$ is used to tune the chemical potential of the semiconductor. 
All measurements are performed at base temperature of a dilution refrigerator.

The experimental setup allows the measurement of the $2\times2$ conductance matrix
\begin{equation}
\label{eq:gmat}
g =
\begin{bmatrix}
    g_{\mathrm{LL}}       & g_{\mathrm{LR}} \\
    g_{\mathrm{RL}}       & g_{\mathrm{RR}} 
\end{bmatrix}
=
\renewcommand\arraystretch{1.15}
\begin{bmatrix}
    \frac{\delta I_L}{\delta V_L}        &- \frac{\delta I_L}{\delta V_R} \\
    -\frac{\delta I_R}{\delta V_L}        & \frac{\delta I_R}{\delta V_R}
\end{bmatrix},
\renewcommand\arraystretch{1}
\end{equation}
where the sign convention is chosen for compatibility with standard two-terminal measurements of $g_\mathrm{LR}$.

Figure~\ref{device}(b) shows $g_{\mathrm{LL}}$ as a function of $V_L$ with $V_R=0$. 
Several peaks occur symmetrically around zero bias. 
We assign the two highest energy peaks to  coherence peaks, a signature of the Bardeen-Cooper-Schrieffer (BCS) density of states in the proximitized semiconductor with an induced gap $\Delta\sim$~250~$\mu$eV, in agreement with previous observation for similar material systems~\cite{chang_hard_2015,vaitiekenas_selective-area-grown_2018}. 
The other peaks are subgap states with energies $E_0 < \Delta$. 
Both the high-bias conductance ($V_L > \Delta$) as well as the subgap peak heights are asymmetric in bias [Fig.~\ref{device}(b)]. 
The nonlocal conductance $g_{\mathrm{LR}}$, measured as function of $V_R$ with $V_L=0$ in Fig.~\ref{device}(c), exhibits features corresponding to the peaks in $g_{\mathrm{LL}}$.
The sign of the peak amplitudes in $g_{\mathrm{RL}}$, however, changes with a sign-change in bias, indicating a strong odd component.
The remaining conductance-matrix elements $g_\mathrm{RR}$ and $g_\mathrm{RL}$ were also measured and exhibit similar features \footnote{See supplement}.

To further explore the symmetry properties, the conductance traces are decomposed into their symmetric,
\begin{equation}
g^{\mathrm{s}}(V_{\mathrm{bias}})=\frac{1}{2}\left(g(V_{\mathrm{bias}})+g(-V_{\mathrm{bias}})\right),
\end{equation}
and antisymmetric,
\begin{equation}
g^{\mathrm{as}}(V_{\mathrm{bias}})=\frac{1}{2}\left(g(V_{\mathrm{bias}})-g(-V_{\mathrm{bias}})\right),
\end{equation}
parts.
Figure~\ref{device}(d) shows $g_{\mathrm{LL}}^{\mathrm{as}}$ as a function of $V_L$, which bears a qualitative resemblance to $g_{\mathrm{LR}}$. 
In fact, $g_{\mathrm{LR}}^{\mathrm{as}}$, the antisymmetric component of $g_{\mathrm{LR}}$, closely matches $g_{\mathrm{LL}}^{\mathrm{as}}$ [dashed line in Fig.~\ref{device}(e)] for low bias, with some quantitative departures observed at high bias.
Reference~\cite{danon_nonlocal_2019} discusses the identified symmetry relation, $g_{\mathrm{LR}}^{\mathrm{as}}=g_{\mathrm{LL}}^{\mathrm{as}}$, as an underlying symmetry of the scattering matrix for $V_{\mathrm{bias}}<\Delta$.
Departures from this symmetry can result from bias-voltage dependence of the effective potentials \cite{buttiker_symmetry_1988}, single-particle scattering into the nominally superconducting lead \cite{danon_nonlocal_2019}, or inelastic processes within the hybrid region \cite{liu_role_2017}.
The fact that departures from symmetry smoothly scale with bias voltage would seem to favor an explanation based on bias-dependent potentials.
In addition, we have found that the observed symmetry departures have relatively little field-dependence, suggesting that they are not due to the presence of dissipation from vortices.

\begin{figure}[t]
    \centering
		\includegraphics[width=1\columnwidth]{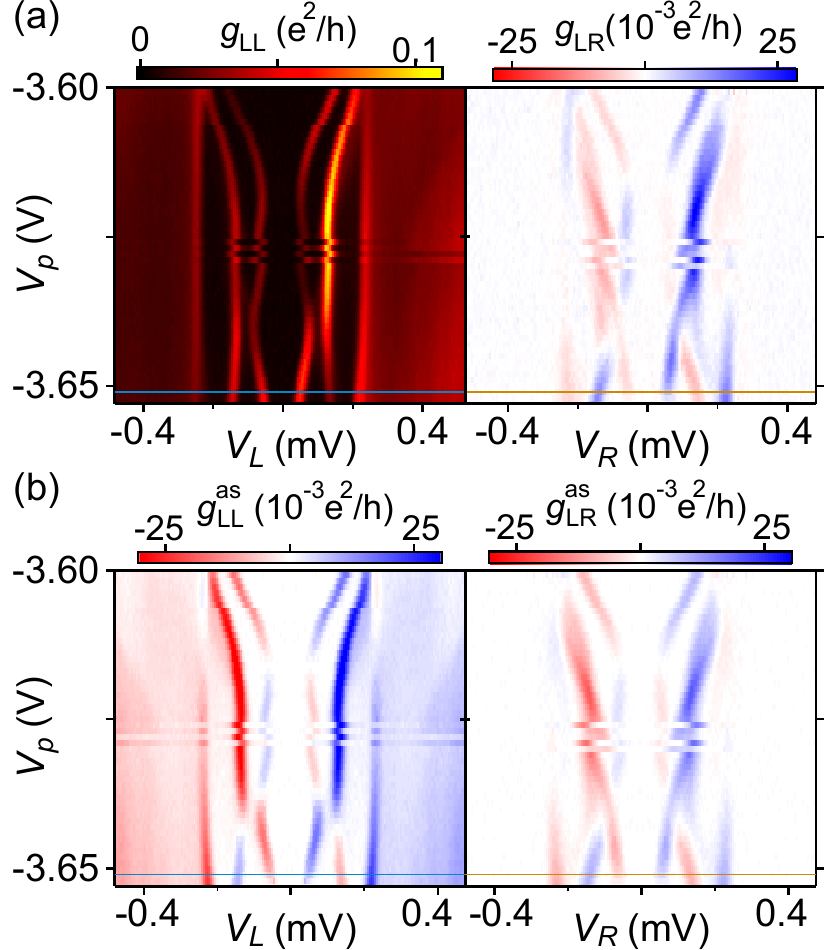}\vspace{-4mm}
    \caption{(a)~Local conductance (left panel) and nonlocal conductance (right panel) as a function of plunger gate $V_p$ and bias voltage. 
    (b) Antisymmetric components of local (left) and nonlocal conductance (right) as a function of plunger gate $V_p$ and bias voltage. 
    The coloured lines indicate the location of the cuts for the local (blue) and nonlocal (orange) traces in Fig.~\ref{device}.}
    \label{twod}
\end{figure}

Next, we investigate the conductance matrix as a function of gate voltage. 
Figure~\ref{twod}(a) shows $g_{\mathrm{LL}}$ ($g_{\mathrm{LR}}$) as a function of $V_p$ and bias $V_L$ ($V_R$). 
$g_{\mathrm{LL}} (V_p)$ makes the assignment of coherence peaks and subgap states more clear. 
The coherence peaks do not move when changing the gate voltage, while subgap states appear at different bias voltages for different $V_p$. 
We attribute these subgap states to Andreev bound states (ABS), although it is not entirely clear where the states are confined. 
We speculate that confinement could result from a Fermi velocity mismatch between the InAs and the Al or disorder in the system.
The ABS subgap features are visible as peaks in the nonlocal conductance $g_{\mathrm{LR}}$ as well. 
Pairs of ABS resonances at positive and negative bias are found to have opposite sign, again indicating a primarily odd functional form.
Symmetry-decomposing the datasets as a function of bias yields $g_{\mathrm{LL}}^{\mathrm{as}}$ and $g_{\mathrm{LR}}^{\mathrm{as}}$ [Fig.~\ref{twod}(b)], which show a remarkable correspondence in general, indicating that the symmetry relationship identified in Fig.~\ref{device} is robust as a function of gate voltage. 
However, in addition to the quantitative high-bias discrepancies already identified in Fig.~\ref{device}(e), there are isolated regions in Fig.~\ref{twod}(b) where $g_{\mathrm{LL}}^{\mathrm{as}}$ and $g_{\mathrm{LR}}^{\mathrm{as}}$ qualitatively differ, associated with crossing of subgap states, e.g. around $V_p\sim~-3.64~\mathrm{V}$; these regions are presently not understood.

It is interesting to note that ABS with the same slope with respect to $V_p$ have the same sign in $g_{\mathrm{LR}}$. 
Further, where the slope changes sign, i.e. at inflection points, the nonlocal conductance disappears and changes sign as well [Fig.~\ref{twod}(a)].
Reference~\cite{danon_nonlocal_2019} shows that, for a spectrally isolated subgap state at energy $E_0$, the sign of the nonlocal conductance at bias voltages near $V_0 = E_0/e$ is generally related to the state's local charge density.
The symmetric part of the conductance, for a bound state coupled to the left(right) leads at rate $\Gamma_{L(R)}$ and energy $|E_0| > \Gamma_{L(R)}, k_B T$ is
\begin{equation}
\label{eq:sym_qlqr}
g_{\mathrm{RL}}^{\mathrm{s}}( V_0 ) = a q_L q_R,
\end{equation}
and the antisymmetric part is
\begin{equation}
\label{eq:asym_qr}
g_{\mathrm{RL}}^{\mathrm{as}}( V_0 ) = a n_L q_R\ \mathrm{sign}( V_0 ).
\end{equation}
where $q_{L(R)} = u_{L(R)}^2 - v_{L(R)}^2$ is the local charge density, $n_{L(R)} = u_{L(R)}^2 + v_{L(R)}^2$ is the local probability density, and $u_{L(R)}$, $v_{L(R)}$ are the left (right) Bogoliubov-de Gennes (BdG) amplitudes.
The general expression for $a$ is cumbersome, but in the limiting case $\Gamma_{L(R)} \gg kT$ it takes the simple form 
$a=\frac{2 e^2}{h}\frac{\Gamma_L\Gamma_R}{(\Gamma_L n_L+\Gamma_R n_R)^2}$. 
Similar expressions also hold for $g_\mathrm{LR}( V_0 )$.
As a consequence, there is a proportionality between the symmetry-decomposed nonlocal conductance and the local charge densities, $g_{\mathrm{RL}}^{\mathrm{s}}( V_0 ) \propto q_L q_R$ and $(g_{\mathrm{LR}}^{\mathrm{as}})\cdot(g_{\mathrm{RL}}^{\mathrm{as}}) \propto q_L q_R$.

Proportionality constants can be eliminated by considering the ratio of conductances, 
\begin{eqnarray}
\label{eq:gratio}
Q_L &\equiv& \frac{g^{\mathrm{s}}_\mathrm{RL} (V_0)}{ g^{\mathrm{as}}_\mathrm{RL}(V_0) } \mathrm{sign}(V_0) \\
\label{eq:qratio}
&=& \frac{u_L^2-v_L^2}{u_L^2+v_L^2}.
\end{eqnarray}
$Q_L$ is a measure of the local charge character; $Q_L=+1$ for a state that is locally electron-like ($u \gg v$), and $Q_L=-1$ for a state that is locally hole-like ($v \gg u$).

\begin{figure}
    \centering
        \includegraphics[width=1\columnwidth]{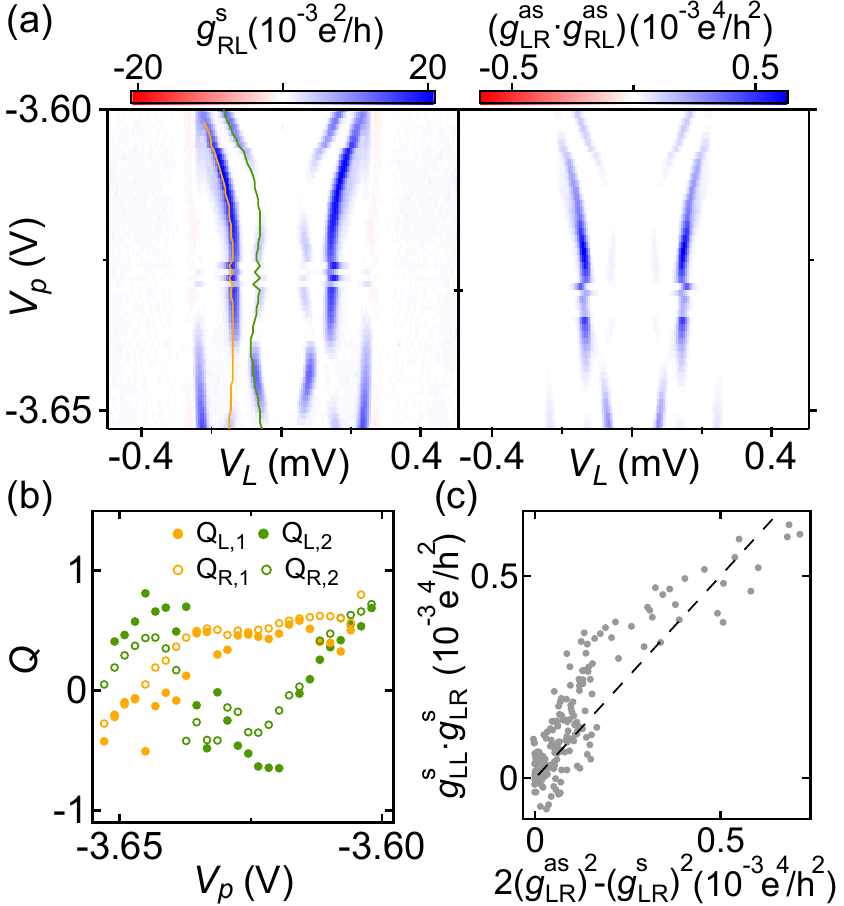}\vspace{-4mm}
    \caption{(a) Symmetric ($g_\mathrm{RL}^\mathrm{s}$ left panel) and the multiplied antisymmetric ($g_\mathrm{LR}^\mathrm{as} \cdot g_\mathrm{RL}^\mathrm{as}$ right panel) components of the non-local conductance as a function of plunger gate voltage and bias. The lines indicate the extracted peak positions for the high (orange) and low (green) energy ABS. 
    (b) Local charge character $Q$ (see Eq.~\ref{eq:gratio}) as a function of $V_p$ for a higher (orange) and lower (green) energy ABS, for both left (full markers) and right (empty markers) side of the device. 
    (c) Left side versus right side of Eq.~(\ref{eq:condRelation}) for all identified subgap peak $V_0$. Dashed line indicates perfect agreement (slope=1, intercept=0). 
    }
    \label{sym}
\end{figure}

Motivated by these theoretical relations, we compare in Fig.~\ref{sym}(a) the measured symmetric component of a single nonlocal conductance, $g_{\mathrm{RL}}^{\mathrm{s}}$, with the product of the antisymmetric components of both nonlocal conductances, $(g_{\mathrm{LR}}^{\mathrm{as}})\cdot(g_{\mathrm{RL}}^{\mathrm{as}})$. 
The plots in Fig.~\ref{sym}(a) are qualitatively similar, as expected from Eqs.~(\ref{eq:sym_qlqr}-\ref{eq:asym_qr}), and suggesting that gate dependence of the BCS charge has a dominant effect on the conductance of subgap peaks.
To further explore the charge of subgap states, the data are analyzed by extracting peak positions of the ABS resonances from $g_{\mathrm{LL}}$.
Decompositions of the nonlocal conductance are evaluated at these positions to obtain $g_\mathrm{RL}^{\mathrm{s}}(V_0)$ and $g_\mathrm{RL}^{\mathrm{as}}(V_0)$. 
$Q_L$ is calculated from the peak values for two different subgap states [orange and green lines in Fig.~\ref{sym}(a)]. 
The local charge character for these states oscillates as a function of $V_p$, highlighting the gate-tunable charge character of bound state [orange and green markers in Fig.~\ref{sym}(b)]. 
Sign changes of $Q_L$ indicate that the state is changing from electron-like to a hole-like, or vice versa. 
We note that the sign changes of $Q_L$ are given by the sign changes of $g_{\mathrm{LR}}^{\mathrm{as}}$. 
Hence the earlier observations that ABS with the same slope with respect to $V_p$ have the same sign in $g_{\mathrm{LR}}$ and change sign at inflection points is entirely consistent with the charge interpretation.
Using ${g_{\mathrm{LR}}}$, $Q_R$ can be analogously calculated [open markers in Fig.~\ref{sym}(b)], and it is found to be similar to $Q_L$ for both bound states. 

The similarity between $Q_L$ and $Q_R$ suggests that the BdG amplitudes have a weak spatial dependence.
In theory, when $u$ and $v$ are spatially uniform Ref.~\cite{danon_nonlocal_2019} predicts that a symmetry relation emerges,
\begin{equation}
    \label{eq:condRelation}
    g_\mathrm{LL}^\mathrm{s}(V_0) \cdot g_\mathrm{LR}^\mathrm{s}(V_0) = 2 [ g_\mathrm{LR}^\mathrm{as}(V_0) ]^2 - [ g_\mathrm{LR}^\mathrm{s}(V_0) ]^2.
\end{equation}
A parametric plot of the left-side versus right-side of Eq.~(\ref{eq:condRelation}) for all identified peaks in the dataset reveals that these quantities are approximately equal [Fig.~3(c)], supporting the view that the coherence factors are spatially uniform.
Indeed, performing a linear fit of the data gives a slope of $0.98 \pm 0.03$ with a small intercept $(0.005)^2 \pm (0.002)^2 (e^2/h)^2$, indicating a good general agreement with the relationship predicted by Eq.~(\ref{eq:condRelation}).
However, there are regions of gate voltage where this analysis systematically fails, such as the previously discussed region $V_P \sim -3.64~\mathrm{V}$, where feature crossings are observed in $g_\mathrm{LR}$.

\begin{figure}[t]
    \centering
       \includegraphics[width=1\columnwidth]{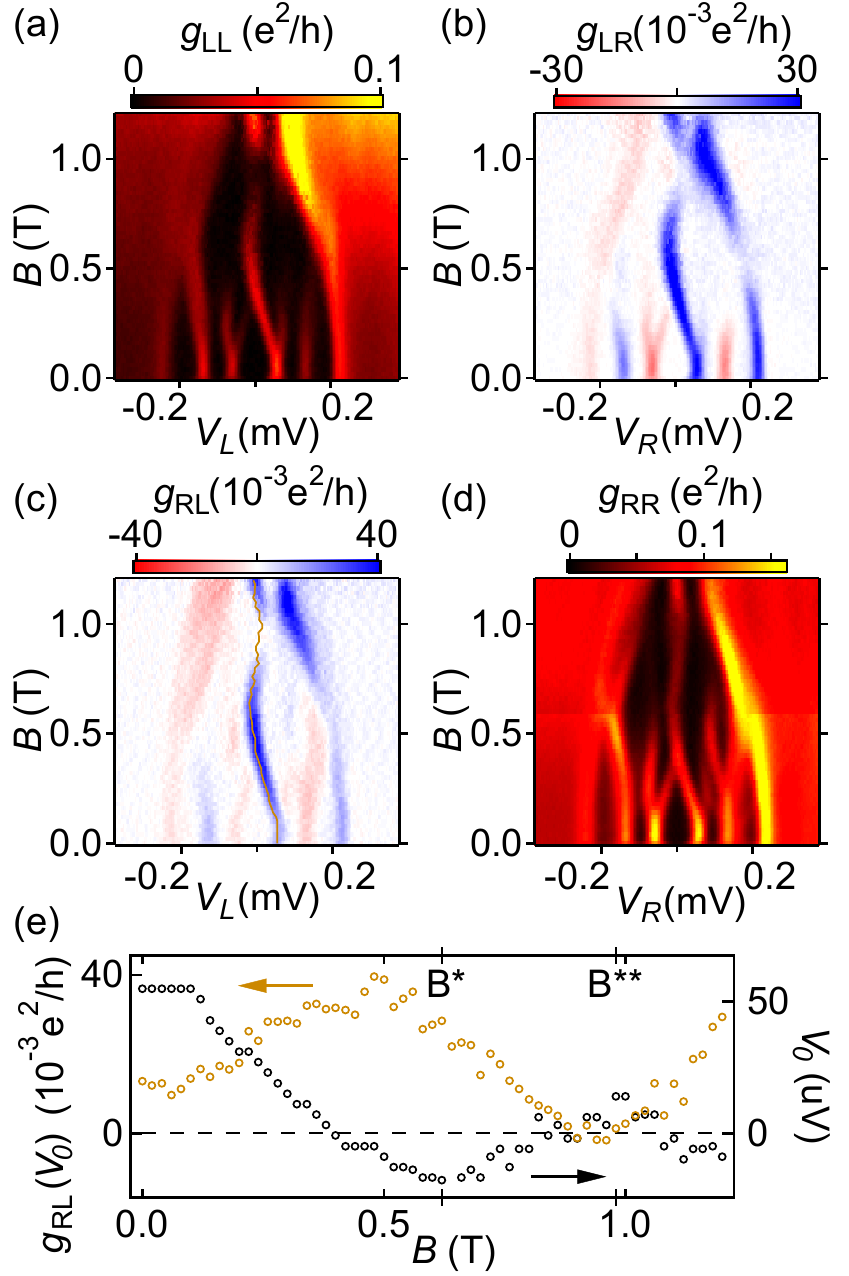}\vspace{-4mm}
    \caption{ Local conductances $g_{\mathrm{LL}}$~(a), $g_{\mathrm{RR}}$~(d) and nonlocal conductances $g_{\mathrm{LR}}$~(b), $g_{\mathrm{RL}}$~(c) as a function of bias voltage and parallel magnetic field $B$. The orange curve overlayed on (c) indicates peaks found from the local conductance $g_{\mathrm{LL}}$. 
    (e) Peak nonlocal conductance $g_{\mathrm{RL}}( V_0 )$ (orange) and corresponding bias voltage, $V_0$, (black) as function of magnetic field $B$.
    Magnetic fields corresponding to the subgap states' energy-turning points, $B^*$ and $B^{**}$, are indicated.
    }
    \label{field}\vspace{-4mm}
\end{figure}

Having established the nonlocal conductance as a tool for characterizing subgap states, we now apply a magnetic field while measuring the conductance matrix.  
Figures~\ref{field}(a) and (d)  show the evolution of $g_{\mathrm{LL}}$ and $g_{\mathrm{RR}}$ in a magnetic field $B$ applied parallel to the wire. 
The ABS evolve as a function of $B$ as detected from both ends of the device. 
As the field initially increases, the low-lying states split and anticross, eventually merging at zero energy.
For further increases in field, the low energy states oscillate around zero energy, a signature typically attributed to hybridized MZM in devices comparable in length to the coherence length~\cite{albrecht_exponential_2016}.
The correlated splitting of zero-bias peaks, measured from both wires ends, was proposed as a ``smoking gun" signature of MZM \cite{dassarma_splitting_2012}, but given the presence of strong correlations at zero magnetic field in this device \cite{anselmetti_end-to-end_2019}, we suggest that this signature is not by itself conclusive.

The nonlocal conductances $g_{\mathrm{LR}}$ and $g_{\mathrm{RL}}$ [Figs.~\ref{field}(b) and (c)] as a function of field have asymmetric features corresponding to subgap states in the local conductance, and also exhibit several changes in sign.
To study the sign of $g_{\mathrm{RL}}$ for the low-energy features, peaks in the local conductance, $g_{\mathrm{LL}}$, are determined [orange line Fig.~\ref{field}(c)], and the peak nonlocal conductance, $g_{\mathrm{RL}}( V_0 )$, is extracted at these points.
Of particular interest is the behavior at the field $B^*\sim0.6$~T, where the energy of the oscillating states has a turning point. 
The nonlocal conductance at $B^*$ is nonzero [Fig.~\ref{field}(e)], inconsistent with the expect behavior for hybridized MZM, which are chargeless at turning points and therefore have vanishing nonlocal conductance \cite{hansen_probing_2018,danon_nonlocal_2019}.
Interestingly, at a higher field $B^{**}\sim1~\mathrm{T}$, there is a second turning point where the nonlocal conductance is small, consistent with the behavior expected for chargeless, hybridized MZM \footnote{At $B^{**}$, $Q$ has large error bars due to small signal, but from Eqs. (\ref{eq:sym_qlqr}-\ref{eq:asym_qr}), $g_\mathrm{RL}(\pm V_0) = 0$ implies $q_R=0$ for nonzero couplings.}.
The non-local transport data therefore reveal the presence of effects outside of the Luchyn-Oreg model \cite{lutchyn_majorana_2010,oreg_helical_2010} at intermediate fields ($B<B^{**}$).
Intermediate-field effects in hybrid nanowires, in particular the possibility of quasi-Majorana modes that emerge before the true topological transition \cite{chun-xiao_andreev_2017,vuik_reproducing_2018,reeg_zero-energy_2018}, have been discussed at length in the literature.
It is, to the best of our knowledge, an open theoretical problem to check if quasi-Majoranas, or other effects \cite{dominguez_zero-energy_2017}, can explain the anomalous charge character that we have inferred near $B^{*}$.

In summary, the observed symmetry relations are consistent with a non-interacting scattering picture, and justify the use of symmetry-decomposed conductance to infer the BCS charge.
At finite field, we have observed correlated splittings of zero-bias peaks, but find that at intermediate fields their charge character is not consistent with a simple Majorana picture.
We anticipate that tunneling spectroscopy at both wire ends, combined with the insights from the nonlocal conductance, will play an important role in substantiating the existence of MZM.

\begin{acknowledgments}
We acknowledge helpful discussions with Jeroen Danon, Roman Lutchyn, Dima Pikulin, and Torsten Karzig. 
This work was supported by Microsoft Project Q and the Danish National Research Foundation. 
C.M.M. acknowledges support from the Villum Foundation. 
\end{acknowledgments}

\end{document}


\title{Supplement to ``Conductance-matrix symmetries of a three terminal hybrid device''}

\author{G.~C.~M\'enard}
\affiliation{Center for Quantum Devices, Niels Bohr Institute and Microsoft Quantum - Copenhagen, University of Copenhagen, Universitetsparken 5, 2100 Copenhagen, Denmark}
\author{G.~L.~Anselmetti}
\affiliation{Center for Quantum Devices, Niels Bohr Institute and Microsoft Quantum - Copenhagen, University of Copenhagen, Universitetsparken 5, 2100 Copenhagen, Denmark}
\author{E.~A.~Martinez}
\affiliation{Center for Quantum Devices, Niels Bohr Institute and Microsoft Quantum - Copenhagen, University of Copenhagen, Universitetsparken 5, 2100 Copenhagen, Denmark}
\author{D.~Puglia}
\affiliation{Center for Quantum Devices, Niels Bohr Institute and Microsoft Quantum - Copenhagen, University of Copenhagen, Universitetsparken 5, 2100 Copenhagen, Denmark}
\author{F.~K.~Malinowski}
\affiliation{Center for Quantum Devices, Niels Bohr Institute and Microsoft Quantum - Copenhagen, University of Copenhagen, Universitetsparken 5, 2100 Copenhagen, Denmark}
\author{J.~S.~Lee}
\affiliation{Department of Electrical Engineering, University of California, Santa Barbara, California 93106, USA}
\author{S.~Choi}
\affiliation{California NanoSystems Institute, University of California, Santa Barbara, California 93106, USA}
\author{M.~Pendharkar}
\affiliation{Department of Electrical Engineering, University of California, Santa Barbara, California 93106, USA}
\author{C.~J.~Palmstr\o{}m}
\affiliation{California NanoSystems Institute, University of California, Santa Barbara, California 93106, USA}
\affiliation{Department of Electrical Engineering, University of California, Santa Barbara, California 93106, USA}
\affiliation{Materials Department, University of California, Santa Barbara, California 93106, USA}
\author{K.~Flensberg}
\affiliation{Center for Quantum Devices, Niels Bohr Institute and Microsoft Quantum - Copenhagen, University of Copenhagen, Universitetsparken 5, 2100 Copenhagen, Denmark}
\author{C.~M.~Marcus}
\affiliation{Center for Quantum Devices, Niels Bohr Institute and Microsoft Quantum - Copenhagen, University of Copenhagen, Universitetsparken 5, 2100 Copenhagen, Denmark}
\author{L.~Casparis}
\email{Equal contribution, lucas.casparis@microsoft.com}
\affiliation{Center for Quantum Devices, Niels Bohr Institute and Microsoft Quantum - Copenhagen, University of Copenhagen, Universitetsparken 5, 2100 Copenhagen, Denmark}
\author{A.~P.~Higginbotham}
\email{Equal contribution, andrew.higginbotham@ist.ac.at}
\affiliation{Center for Quantum Devices, Niels Bohr Institute and Microsoft Quantum - Copenhagen, University of Copenhagen, Universitetsparken 5, 2100 Copenhagen, Denmark}

\maketitle

\section{Conductance matrix measurement}
Each element of the conductance matrix,
\begin{equation}
\label{eq:gmat}
g =
\begin{bmatrix}
    g_{\mathrm{LL}}       & g_{\mathrm{LR}} \\
    g_{\mathrm{RL}}       & g_{\mathrm{RR}} 
\end{bmatrix}
=
\renewcommand\arraystretch{1.15}
\begin{bmatrix}
    \frac{\delta I_L}{\delta V_L}        &- \frac{\delta I_L}{\delta V_R} \\
    -\frac{\delta I_R}{\delta V_L}        & \frac{\delta I_R}{\delta V_R}
\end{bmatrix},
\renewcommand\arraystretch{1}
\end{equation}
is in general a function of both DC biases $V_L, V_R$.
We experimentally observe that the left column of $g$ depends only on $V_L$ and the right column only on $V_R$, as expected in the absence of bias-dependent effective potentials \cite{danon_nonlocal_2019}.
In general, the conductance matrix of a three-terminal device, $g_{\alpha \beta}=d I_\alpha / d V_\beta$, is a $3\times3$ matrix.
However, conservation of current and insensitivity to overall offset voltages imply the sum rules $\Sigma_\alpha g_{\alpha \beta} = \Sigma_\beta g_{\alpha \beta} = 0$, which allow the remaining row and column of the $3 \times 3$ matrix to be inferred from the quantities in Eq.~(\ref{eq:gmat}).
The sum rules are valid when there are no spurious paths to ground, which we have experimentally verified to exceed $1~\mathrm{G\Omega}$ in our setup.

\begin{figure}[h]
\includegraphics[scale=1]{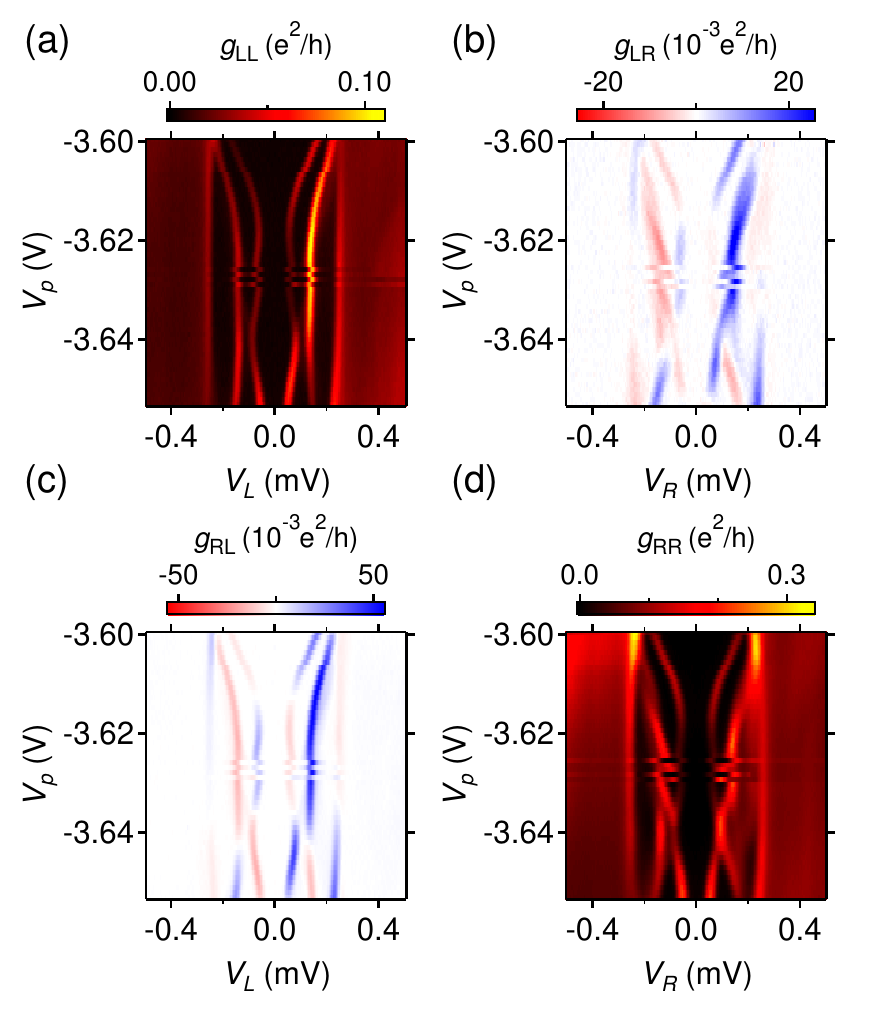}
\caption{\label{fig:S1} Conductance matrix measurment a zero magnetic field. Local conductances $g_{\mathrm{LL}}$~(a), $g_{\mathrm{RR}}$~(d) and nonlocal conductances $g_{\mathrm{LR}}$~(b), $g_{\mathrm{RL}}$~(c) as a function of bias voltage and plunger gate $V_p$.
}
\end{figure}

For completeness, Fig.~\ref{fig:S1} shows the measurement of all conductance-matrix entries as a function of bias voltages $V_L$, $V_R$ and plunger gate $V_p$. 
We note that the data is taken by sweeping the left bias $V_L$, while $V_R$=0 and then $V_R$ is swept, while $V_L$=0, only then the gate voltage $V_p$ is changed.

\section{Correspondence between ABS response to gate and $Q$}

\begin{figure}[h]
\includegraphics[scale=1]{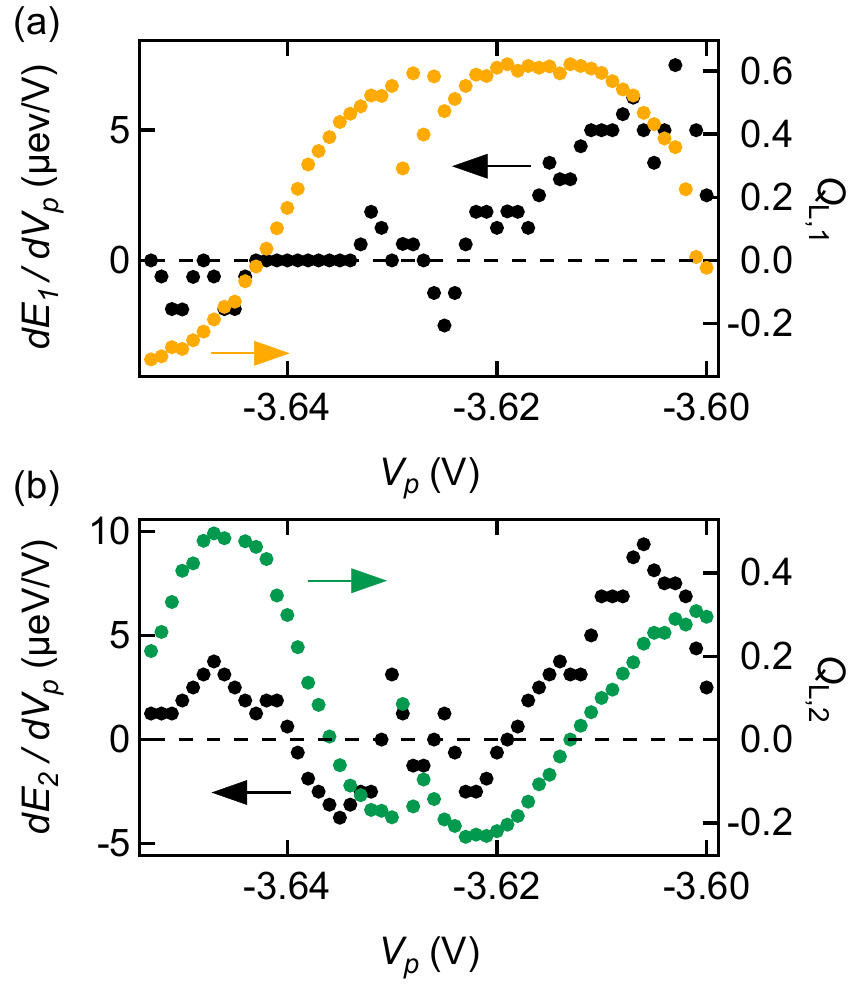}
\caption{\label{fig:S2} Response to plunger gate voltage and extracted charge.(a) The derivative of the energy of the higher energy ABS $E_1$ with respect to plunger gate voltage $V_p$ $dE_1/dV_p$ as a function of $V_p$ on the left axis (black). On the right axis the extracted $Q_{\mathrm{L,1}}$ from non local measurements (orange) for the higher energy state is plotted as a function of $V_p$.(b) The derivative of the state energy of the lower energy ABSe $E_2$ with respect to plunger gate voltage $V_p$ $dE_2/dV_p$ as a function of $V_p$ on the left axis (black). On the right axis the extracted $Q_{\mathrm{L,2}}$ from non local measurements (green) for the lower energy state is plotted as a function of $V_p$.
}
\end{figure}

As described in the main text, the quantity $Q_L=(u_L^2-v_L^2)/(u_L^2+v_L^2)$ measures the local charge character of the ABS. 
It is natural to ask if the gate-voltage response of the ABS correlates with the extracted charge for a given ABS, as one would expect from theory \cite{danon_nonlocal_2019}.
To investigate, the numerical derivative of the energy of the ABS $E_{1}$ with respect to plunger gate $V_p$ is plotted for the higher energy ABS as a function $V_p$ [black in Fig.~\ref{fig:S2}(a)]. 
On the right axis of Fig.~\ref{fig:S2}(a), $Q_{\mathrm{L,1}}$ is plotted as a function of $V_p$ (orange). 
There does not seem to be a correlation between the two quantities. 
This can also be seen from the raw data in Fig.~\ref{fig:S1}(c) where the higher energy ABS is almost not moving in energy, but still has a sizeable nonlocal signal.
The clear discrepancies between extracted local charge character and the response to plunger gate for this state are not understood.
Discrepancies could arise if the gate voltage couples to device parameters other than the chemical potential, such as the spin-orbit coupling or $g$-factor.
Deviations could be enhanced by the high-bias departures from the symmetry relations identified in the main text Fig.~1(e).

Moving to the lower energy ABS with energy $E_2$, Fig.~\ref{fig:S2} shows $dE_{2}/dV_p$ as a function of $V_p$ (black). On the right axis of Fig.~\ref{fig:S2}(b) $Q_{\mathrm{L,2}}$ is plotted as a function of $V_p$ (green). 
One can recognize a correlation between the two quantities, although $dE_{2}/dV_p$ and $Q_{\mathrm{L,2}}$ appear to be shifted by roughly 50~mV in $V_p$. 
Looking at the raw data in Fig.~\ref{fig:S1}(c) one can see that indeed the disappearance of the nonlocal signal for the lower energy state and thus $Q_{\mathrm{L,2}}=0$ appear before the actual inflection points of the state energy.
Similar behavior was observed in numerical simulations, and was attributed to the fact that the inflection points are determined by the global charge, whereas the nonlocal conductance ratio measures the local charge \cite{danon_nonlocal_2019}.